\documentclass[
  journal=largetwo,
  manuscript=article-type,
  year=2020,
  volume=37,
]{cup-journal}

\usepackage{amsmath}
\usepackage[nopatch]{microtype}
\usepackage{booktabs}

\title{Converting the ANU 2.3 Telescope to Fully Automated Operation}

\author{I. Price}
\affiliation{Research School of Astronomy and Astrophysics,\\
  Australian National University,\\
  Canberra, ACT 2611, Australia}
\email[I. Price]{ian.price@anu.edu.au}

\author{J. Nielsen}
\affiliation{Research School of Astronomy and Astrophysics,\\
  Australian National University,\\
  Canberra, ACT 2611, Australia}

\author{C. Lidman}
\affiliation{Research School of Astronomy and Astrophysics,\\
  Australian National University,\\
  Canberra, ACT 2611, Australia}

\author{J. Soon}
\affiliation{Research School of Astronomy and Astrophysics,\\
  Australian National University,\\
  Canberra, ACT 2611, Australia}

\author{T. Travouillon}
\affiliation{Research School of Astronomy and Astrophysics,\\
  Australian National University,\\
  Canberra, ACT 2611, Australia}

\author{R. Sharp}
\affiliation{Research School of Astronomy and Astrophysics,\\
  Australian National University,\\
  Canberra, ACT 2611, Australia}


\begin{document}

\begin{abstract}
  The operation of the ANU 2.3\,m telescope transitioned from classically scheduled
  remote observing to fully autonomous queue scheduled observing in March 2023.
  The instrument currently supported is WiFeS, a visible-light low-resolution
  image-slicing integral field spectrograph with a $25''\times38''$ field of view (offering precision spectrophotometry free from aperture effects).
  It is highly suitable for rapid spectroscopic follow-up of astronomical transient events and regular cadence  observations.
  The new control system implements flexible queue scheduling and supports
  rapid response override for Target-of-Opportunity observations. The ANU 2.3\,m is the
  largest optical telescope to have been retro-fitted for autonomous operation to date,
  and it remains a national facility servicing a broad range of science cases. We
  present an overview of the automated control system and report on the first six months of
  continuous operation.
 \end{abstract}

\section{Introduction} \label{sec:intro}

The 2.3\,m telescope at Siding Spring Observatory is owned and operated by the
Australian National University. It was designed to be operated by a single
observer from the control room located in the co-rotating telescope enclosure.
Low-cost high-bandwidth networking enabled remote observing, but this was
only partially realized until the deployment of the Wide Field Spectrograph (WiFeS),
as the instrument configuration and the calibration subsystems were entirely
under software control \citep{DopitaWiFeS}.

The growth of transient astronomy and the accompanying all-sky transient surveys has transformed the way that telescopes observe the night sky. These surveys can provide hundreds to thousands of candidates with an unknown astrophysical origin per night and the increasing number and interests in the short-term evolution of these phenomena has resulted in an increased demand for
Target-of-Opportunity (ToO) override. This demand is expected to further increase
as new facilities that detect transients start operations (Rubin\footnote{Vera C. Rubin Observatory}, CTAO\footnote{Cherenkov Telescope Array Observatory}, DREAMS\footnote{Dynamic REd All-sky Monitoring Survey}).
Remote operation made automated operation feasible and the predicted demand for
rapid follow-up of astronomical transients made it highly desirable.

To support automated operation, the software control system of the telescope was
upgraded and the software control system of the WiFeS instrument was completely
redesigned and replaced. However, no significant changes were made to the control
system hardware. In section~\ref{sec:ctrlsys} we outline core elements of the
new control system. Section~\ref{sec:initialops} presents a summary from the
first half-year of operation. In section~\ref{sec:discuss} we discuss the
change in demand for the facility over this relatively short period, and the
role realistic simulators played in a low-cost software refurbishment project
that is increasing the scientific output from a mature facility.

\section{Automated Observatory Control System} \label{sec:ctrlsys}

The key design requirements for the automated system were support for the established
modes of operation of the WiFeS instrument, near-instantaneous
ToO override, autonomous queue scheduling, and robust self preservation. The minimum
unit of work from the perspective of the astronomer, and the perspective of
the control system, are fundamental elements of the final design. We chose
a single instrument setup and exposure configuration of a pre-defined observing
mode as the unit of work for the latter, and an ordered sequence of these
\emph{observation blocks} as the unit of work for an observer. This decomposition
is similar to the European Southern Observatory model for queue scheduled
observations on the Very Large Telescope \citep{10.1117/12.316474}, and the
\emph{observation block} as the quantum of work has remained a stable element
in their system throughout its evolution \citep{2018Msngr.171....8H}.
In the aperture class similar to the ANU 2.3m,
the fully robotic Liverpool Telescope has a different approach. Their system manages programs of
observations, and the observer has the option of defining their own observing sequence \citep{2010SPIE.7737E..11S}. Our approach knowingly traded complete flexibility for
reduced complexity in the pursuit of robustness.

\begin{figure}[hbt!]
\centering
\includegraphics[width=1.0\linewidth]{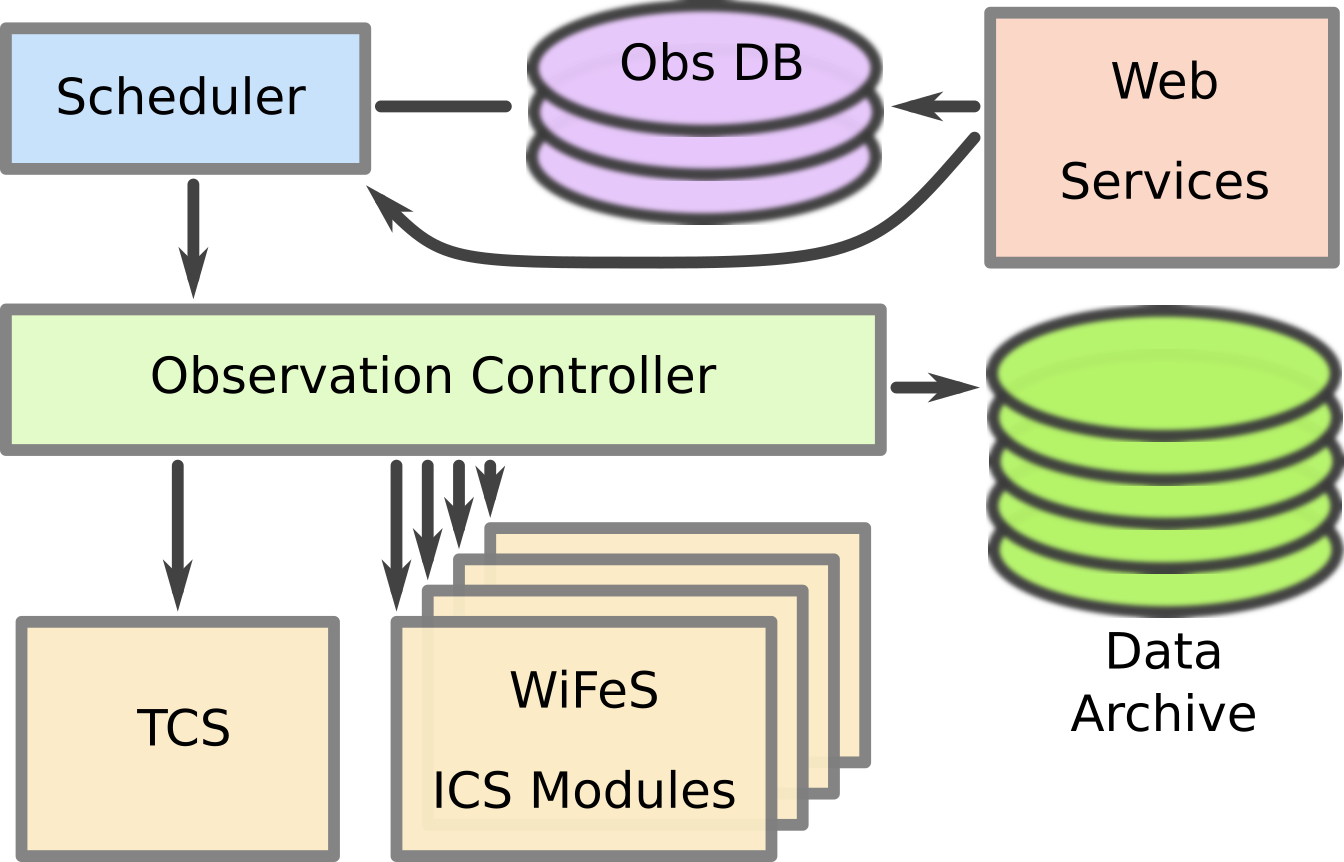}
\caption{Schematic of the high-level architecture of the automated observatory control system.}
\label{fig:archoverview}
\end{figure}

The high-level architecture of the control system is shown in Figure~\ref{fig:archoverview}.
The instrument control system (ICS) is modular and distributed. It's a collection of subsystems
that includes the Telescope Control System (TCS), CCD detector controllers, industrial motion
controllers, and the acquisition camera. They are independent, can function in isolation, and there's no direct channel of communication between them. The architecture supports multiple instruments and all future
instruments will be required to operate within this general scheme of control.

The Observation Controller process consumes observation blocks and coordinates all of the
activities of the ICS components. Operating modes of the instrument are implemented
in the business logic of this process, not in the ICS layer below. The Observation
Controller is directed by the Scheduler. All decision making is implemented by the
Scheduler, including assessment of the current observing conditions, observation selection,
override for a ToO, or suspension of observing due to poor weather conditions. Universal
coordinated time, meteorological data, and the pool of observations stored in a database
underpin the operation of the Scheduler.

All processes in the system use a request-response client-server model for communication.
The ICS modules report status information on request. The Observation Controller process uses a
query-command-verify strategy in the implementation of all operational sequences. An initial
state is never assumed and a specific final state is not required.
The command-verify steps are skipped if the initial query indicates no action is
necessary to progress through the sequence. Each ICS component provides a relatively
thin layer of abstraction around the hardware it controls. That hardware does not need to
have a native asynchronous control interface but the ICS software process must. This
ensures status requests are serviced promptly and the communication channel is
not blocked. The general approach is simple and it can be formed around virtually any
hardware device. It is highly suitable for retro-fitting to existing instruments with
bespoke hardware and modest data rates. We used a consistent communication system for
all ICS modules, but note that it was not strictly necessary to do so.

Using configurable, but pre-defined observing modes, ensures the system is operated
safely and efficiently. Actions are executed in parallel where possible, and in serial
when they are not. Sensibly ordering the operations to minimize overheads requires
a detailed understanding of the nuances of subservient ICS components. When exploring
use-cases it became clear that only a small number of very experienced 2.3m observers
had knowledge of the system sufficiently detailed to maximize time-efficiency when
observing. Offering only pre-defined operating modes leveraged both the practices of the
most experienced observers and the detailed understanding of various subsystems held by the engineering team. It traded efficiency for flexibility, but all known modes of operation were implemented
and the set of modes is readily extendable. Pre-defined operating modes also gave the
development team full control of the cancellation points that are required for
interrupting an observation and swiftly commencing the execution of a different
sequence. Robust interruption is necessary for rapid response to ToO requests and
critical for self preservation in the event of bad weather.

\subsection{Observations} \label{observations}

\begin{figure*}[htb!]
\includegraphics[width=0.9\linewidth]{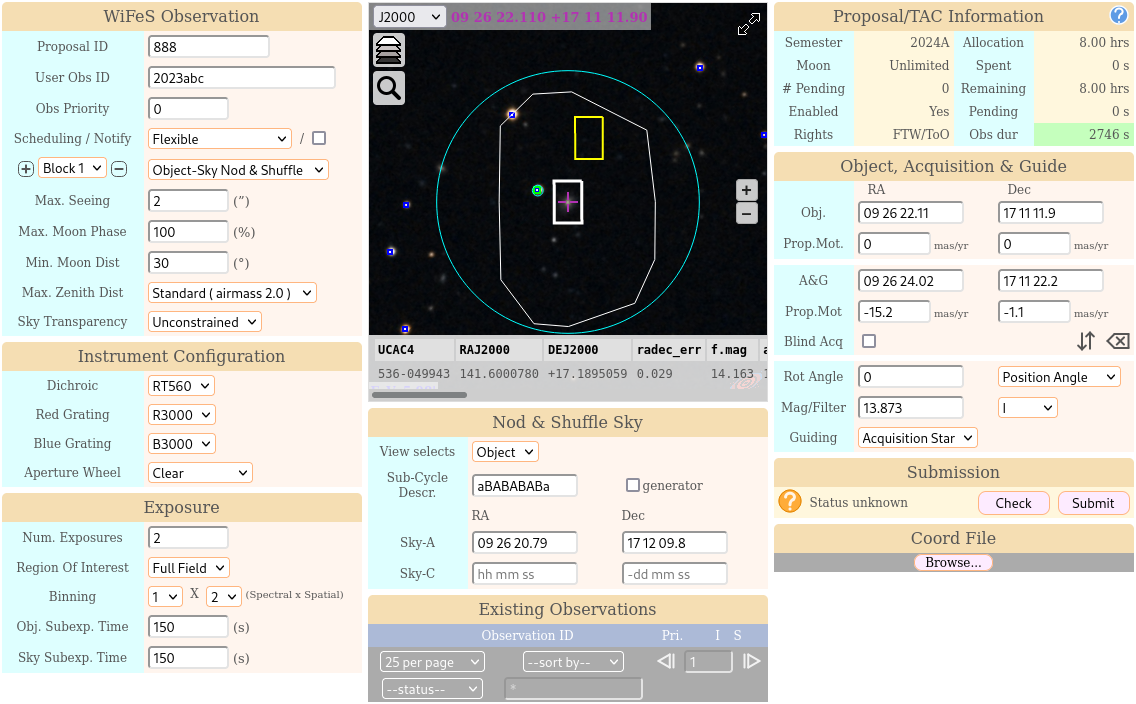}
\caption{A screenshot of the observation preparation tool with a complete nod-and-shuffle observation request.} 
\label{fig:webtool}
\end{figure*}

Astronomers granted time on the telescope prepare their observations with a web-based
tool that incorporates the Aladin Lite v2 sky viewer \citep{2014ASPC..485..277B}.  This facilitates construction of observation requests, comprised of observation
blocks and observing constraints, and provides a mechanism
for adding them into the pool of possible observations. Figure~\ref{fig:webtool} shows the tool
during the preparation of a nod-and-shuffle observation for transient follow-up. The observation request includes
parameters necessary for scheduling and an ordered list of observation blocks.
The observation blocks are passed as-is to the Observation Controller at the point of
execution and they are opaque to the Scheduler. Observation blocks are decoded in a hierarchical
process that progresses towards more specific detail about the sequence to be executed.
Being opaque to all other parts of the system enabled the control system to be developed
and tested with just a couple of operating modes, and expanded later with development
localized to the Observation Controller and the validation of submitted observations. 

\subsection{Observation Selection} \label{obsselect}

The Scheduler partitions each 24--hour period into eight sessions. The boundaries between sessions
are either fixed in local-time or relative to twilight. End-user observations are allotted only
to the session between astronomical twilight. The system uses some sessions for calibration observations
on behalf of all users, and others are reserved for use by technical
staff maintaining the facility. Observations allotted to a session are drafted into either the
flexible queue or the ToO queue. The ToO queue is serviced if non-empty and an executing non-ToO observation
will be terminated when a ToO observation can be serviced. Both queues are managed as priority queues using a rank
metric. At the point of observation selection, the Scheduler recomputes the rank metric for each observation
to take the current observing conditions, constraints, and other time dependent factors into account.
The highest ranked observation is selected and execution of the list of observation blocks is delegated to
the Observation Controller.

The rank metric, $m$, is a complex multivariate function
that, broadly, implements the policies of the time allocation committee. It is structured as a
weighted linear sum of functions that return values bound to the interval $[-1,1]$,

\begin{equation}
    m = \sum_{i} f_{i}(\bar{x})w_{i}
\end{equation}

where $f_{i}$ is a policy function, $\bar{x}$ is the vector of current conditions and observation parameters, and $w_{i}$ is the weight applied to a specific policy. Each function implements one aspect of the policy, with positive values a bias towards selection and negative values biasing against selection. There are policy functions for favouring observing
\begin{itemize}
    \item as targets transit the meridian
    \item within a specific range of seeing
    \item within a specific range of lunar phase
    \item at larger angular distances from the Moon
    \item if the time allocation has not been consumed
    \item according to scientific merit
\end{itemize}

Although each element of the policy has a role to play, the largest weights are applied to scientific merit and observing at small hour angles. To avoid having high science value observations done at large hour angles the science merit policy function has a strong negative bias at $|HA|>1.5$ hours.
Our metric is an alternative form of the Liverpool Telescope's original rank metric \citep{10.1117/12.278824} that does not formally separate the efficiency functions from the fairness function. The decomposition and bound range of each policy function is merely for convenience as it makes adjusting the relative weights of the policy elements reasonably intuitive.

\subsection{Acquisition}

\begin{figure}[htb!]
\includegraphics[width=1.0\linewidth]{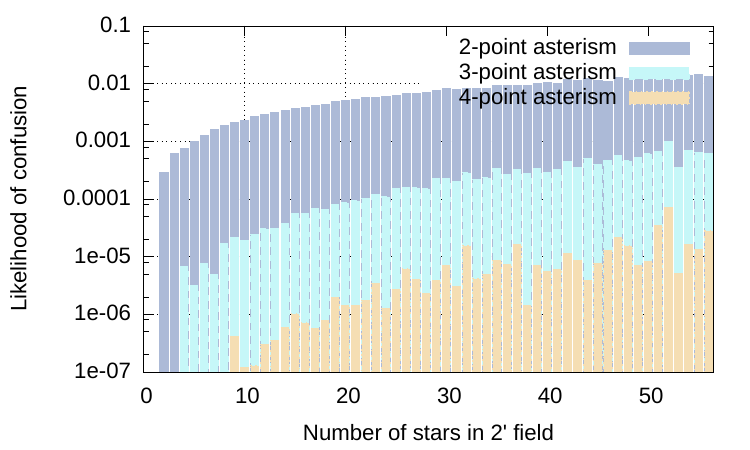}
\caption{The likelihood of asterism mis-identification with 2, 3 \& 4 star asterisms as a function of field density in the WiFeS acquisition camera. All possible asterisms were considered from 250000 randomly selected field centers using the UCAC4 catalog.} 
\label{fig:confusion}
\end{figure}

The RMS pointing error of the 2.3m is around 5 arcsec. Some science cases are tolerant of this
blind pointing error owing to the $25\times38''$ field of WiFeS, but we have a goal of acquiring
with an error of less than 0.5''.
The field of the WiFeS acquisition and guide camera is approximately 3 arcmin across
but is strongly vignetted at the edge and the center of the field. Central vignetting is due to the
hole in the reflector that passes the science field to the integral field unit. A two-stage process
is used to refine the pointing correction during acquisition. In the first instance
the brightest point sources are extracted from the full field image \citep{1987PASP...99..191S} and compared with a
source list extracted from a star catalog. The platescale and orientation of the field are accurately
known, so the plate solver algorithm only searches for the translation required to best align the extracted
sources with the catalog. The field is small and irregular, and the exposure depth is adaptive. It
is not unusual to extract fewer than three stars. A Monte-Carlo analysis showed that pointing
errors as large as 60 arcseconds could be unambiguously resolved in over 99.9\% of cases with
only two stars detected. Figure~\ref{fig:confusion} shows the likelihood of confusion for two, three and four star asterisms
as a function of the number of stars in a 2' field, with a 1.5'' alignment tolerance. The likelihood
of confusion increases with field density, but so too does the number of stars that can be extracted
from the image to form the asterism. General plate solvers such as Astrometry.net \citep{2010AJ....139.1782L}
use geometric hashes of three and four star asterisms and require multiple matches. Our approach is fast, does not
require pre-processing of a star catalog, incorporates proper motion, and can be applied to small
low density fields with irregular vignetting. It reliably resolves the issue of source ambiguity
in both sparse and dense fields and largely removes blind pointing error in one iteration. The second stage
is a classic iterative correction based on a subimage of the user-chosen acquisition star.

\subsection{Auto-recovery}

In the automated control system, the Scheduler replaces the human operator. Experienced observers
were familiar with warning and error messages as well as the procedures for resetting and recovery
from the more common transient hardware faults. Our initial design therefore mapped that experience into
technical observation modes instigated by the Scheduler. During on-site testing this proved less
successful than expected due to contention between the various layers of the control system. It
became apparent the telescope control system was better placed to implement auto-recovery actions,
and merely inform the higher layers that it was busy recovering via its status messages. This
approach increased the separation of responsibility of the various components and resulted in
simpler implementations of both the telescope control system and the observation controller.
Auto-recovery was not implemented in the WiFeS instrument controllers because transient faults
are very rare. 

\section{Initial Six Months of Operation} \label{sec:initialops}

\begin{figure}[htb!]
\includegraphics[width=1.0\linewidth]{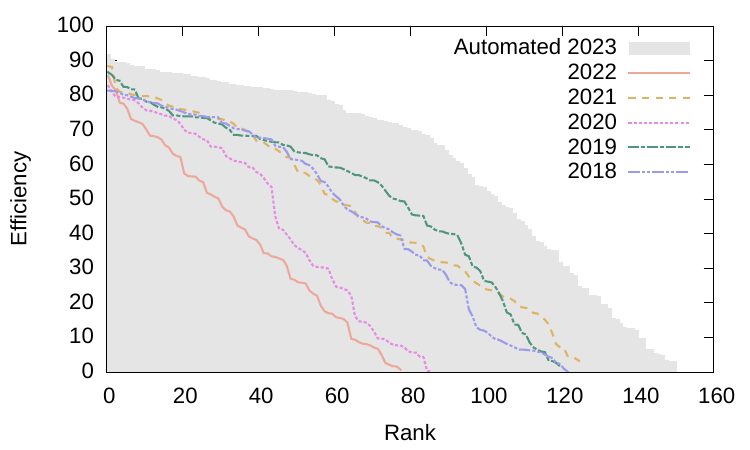}
\caption{Observing efficiency of each night between March 22 and September 21, ranked in order of decreasing efficiency. The automated system produced the most efficient night and consistently outperforms a human observer. The poor apparent performance in 2022 is largely due to highly under-subscribed bright time and in 2020 operation of the telescope was suspended during COVID lockdown.}
\label{fig:efficiency}
\end{figure}

The Automated Observatory Control System for the 2.3m telescope and WiFeS instrument has been in
continuous operation since $11^{th}$ March 2023. In the six months between the vernal and autumnal equinoxes
of 2023 a total of 3377 science observations were successfully completed, with 13 of these scheduled as Target-of-Opportunity observations. The total time on target
was 943 hours. The efficiency of every night in this date-range in recent years was computed from
the data in the archive. It is plotted against the efficiency rank in Figure~\ref{fig:efficiency}, where
efficiency is defined as the ratio of total time observing targets to the time between astronomical twilight.
Slewing between targets is not counted as observing time.
Low efficiency nights are generally the result of poor weather, and an observer and the automated system
both have a subjective definition of poor conditions, so variation in the tail of the distribution is
expected. However at the high efficiency end of the range the nights are completely clear. This shows
the automated system consistently outperforms a human observer in time-utilization of the facility.

\begin{figure}[htb!]
\includegraphics[width=1.0\linewidth]{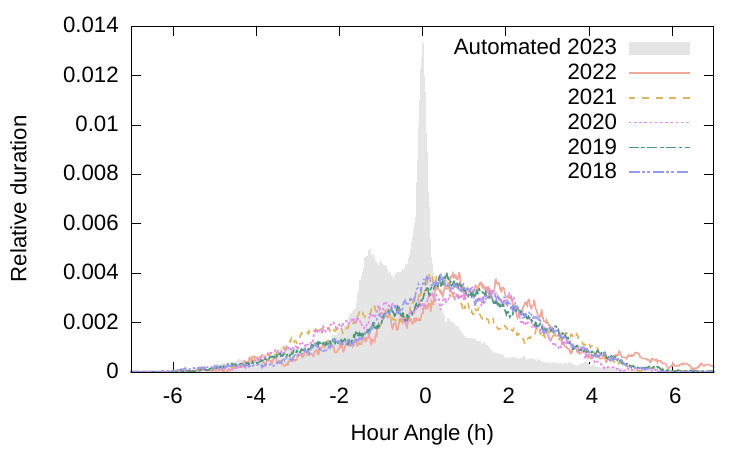}
\caption{Normalized distribution of time on target with hour-angle. Automated scheduling results in more
time spent observing near the meridian and to the East as high priority targets rise.
The data from the five years prior to automated operation indicate human
observers bias towards observing objects in the West.} 
\label{fig:hahist}
\end{figure}

The normalized distribution of time spent observing at a given hour-angle is shown in
Figure~\ref{fig:hahist}. This clearly shows the effect of the scheduling rank to favour observing
as objects cross the meridian. The scheduling rank metric supports a reduced range of preferred hour-angle
for high priority observations. This feature produces the secondary peak at -1.5h, as limited competition amongst
high priority observations leads to their selection as they rise. It also shows that classical
scheduling leads to a broader distribution with observers marginally favouring observing to the West. The broader distribution implies observing at a higher airmass on average. This is expected given the limited time window of a classically scheduled observing run, but it is also undesirable. We have no reason to suppose the slight preference by observers for positive hour angles is either intentional or advantageous and have not sought to replicate this biased behaviour via the rank metric.
  
\section{Discussion} \label{sec:discuss}

In the years prior to automated operation the 2.3m telescope was slightly under subscribed. After
six months of operation the demand has increased significantly, with dark time now over subscribed
by a factor of three. Requests for ToO observations have also increased. We attribute the increase in
demand to a significant reduction in the burden of observing and the increased return on investment.
Automated operation amortizes the risk of downtime and poor weather across all observing programs,
so effort spent preparing observations is highly likely to be rewarded with data. Users do not need
to alter their sleep patterns or plan around other time commitments. The observations are executed
efficiently and consistently, with time-critical and unconstrained observations seamlessly interleaved.
Although ToO observations were previously supported, the observation were made by the scheduled
observer at a time they found convenient. As not all observers were expert using all operating
modes this resulted in mixed data quality and an uncertain delay in execution of the observations.
Automation resolves both of these issues.

The system was designed to continue to support a broad range of science programs. All
of the operating modes that were available when observing remotely are supported, and preserving that
functionality retained the established user community. Offering small allocations and
promising a high likelihood of obtaining data facilitates test-case observations, which in
turn is growing the user community.

The ANU 2.3m is the largest optical telescope that has been retrofitted for completely autonomous
operation. Although no significant changes were made to the hardware, the software control system
is a near-complete replacement. Removing all legacy software eliminated stale and deprecated
technologies. It also simplified many systems, largely due to the absence of requirements to support
user interaction and responsive visualization.

The project duration was 36 months and most of the software development was done in
the first two years. Testing with the deployed hardware was contained to three observing runs between
months 11 and 24. Because access to the hardware was extremely limited
a simulator for every component of the ICS was also developed in lock-step with their
hardware interfacing counterpart. These had precisely matched interfaces
and semi-realistic timing for their state changes. The hardware interfacing ICS modules are independent,
but this is not the case for all of the simulators. The simulators for the telescope control system
and the acquisition camera were weakly coupled. This allowed the simulated images produced by the
acquisition camera to realistically model the field the telescope was pointing to. This level of
detail in the ICS simulators was critical to development of the acquisition and guiding algorithms,
and the more complex observing modes such as Nod-and-Shuffle.
The need for this level of complexity in the simulators was identified after the first observing run 11
months into the project.
Having complete and semi-realistic simulators for the ICS ensured the operating sequences implemented
by the Observation Controller would execute successfully. This also enabled use of the Scheduler and
extensive end-to-end testing in complete isolation from the hardware. By investing time and effort into
developing these ICS module simulators we ensured the effective use of the limited time available for
on-site testing with the real hardware. In total, 14 nights spread over three observing
runs were used, and automated operation was demonstrated during the very first observing run.
Software development reduced to maintenance activities during the final 12 months, and two more week-long observing runs were
scheduled for science verification with the participation of the existing user community. Minor issues
arising during the period of science verification and maintenance were addressed via simulation prior to commissioning and the
permanent transition to fully automated operation.
Inflation corrected, the cost of developing the entire automated control system was about one third
the cost of the software development effort for the original WiFeS control system and the remote observing
system that it has now replaced.

\section{Conclusions} \label{sec:conc}

The ANU 2.3m telescope has been successfully converted to fully autonomous operation. The comprehensive
software-only upgrade of the control system supports the continued operation of WiFeS as an open-access
general purpose instrument. The new control system enables rapid-response target-of-opportunity observations,
ensuring the user community is well placed to leverage the anticipated increase in the detection rate and number of transient objects. The first six months of operation has demonstrated an increase in operating efficiency
and seen growth in demand. In an era where 2m-class telescopes are facing closure this project is expected
to extend the operating lifetime of the 2.3m telescope by at least ten years.

\begin{acknowledgement}
We thank Robert J. Smith for generously offering advice on a broad range of
topics based on decades of experience with the robotic Liverpool Telescope.
\end{acknowledgement}

\paragraph{Funding Statement}
The automation of the 2.3-metre telescope was made possible through funding
provided by the Centre of Gravitational Astrophysics at the Australian
National University.

\printbibliography

\end{document}